\documentclass[12pt]{article}
\usepackage{amsmath}
\usepackage{amssymb}
\usepackage{mathrsfs}
\usepackage[T1]{fontenc}
\usepackage{amsthm}
\usepackage{enumerate}
\def\be{\begin{equation}}
\def\ee{\end{equation}}
\newcommand{\eeq}[1]{\label{#1}\end{equation}}
\def\bea{\begin{eqnarray}}
\def\eea{\end{eqnarray}}

\def\coor{coordinate}
\def\eqn{equation}
\def\tfn{transformation}
\def\fn{function}
\def\cond{condition}

\def\+{{+\!\!\!+}}
\def\-1{^{-1}}

\def\half{\frac{1}{2}}
\def\unit{{\bf 1}}

\def\e{{\rm e}}

\def\real{{\mathbb R}}

\def\sm{$\sigma$--model}

\def\pltp{Poisson--Lie T--pluralit}
\def\pltd{Poisson--Lie T--dualit}
\def\dd{Drinfel'd double}

\def\text{}
\def\wt{\widetilde}

\def\wh{\widehat}
\def\dotplus{\stackrel{{\bf.}}{+}}

\def\cf{{\mathcal {F}}}

\def\cd{{\mathfrak d}}

\def\cg{{\mathfrak g}}

\def\htil{\hat{h}}
\def\gtil{\tilde{g}}
\def\gbar{\bar{g}}
\def\xitil{\tilde{\xi}}

\def\ghat{\hat{g}}

\begin{document}

\begin{center}
{\large \bf{On Uniqueness of T--duality with Spectators}}

\vskip .2in

Ladislav Hlavat\'y, Filip Petr\'asek\\
{\em Faculty of Nuclear Sciences and Physical Engineering, \\
Czech Technical University in Prague,\\
B\v rehov\'a 7, 115 19 Prague 1, Czech Republic}
%\vskip .05in E-mail: Ladislav.Hlavaty@fjfi.cvut.cz, Ivo.Petr@fjfi.cvut.cz, stepavoj@fjfi.cvut.cz}

\end{center}

\vskip .4in

\abstract{We investigate the dependence of non-Abelian T--duality on
various identification of the isometry group of target space with its orbits, i.e. with respect to the
location of the group unit on manifolds invariant under the isometry
group. We show that T--duals constructed by isometry groups of
dimension less than the dimension of the (pseudo)-Riemannian manifold
may depend not only on the initial metric but also on the choice of
manifolds defining positions of group units on each of the
sub-manifold invariant under the isometry group. We investigate
whether this dependence can be compensated by coordinate
transformation.}

\section{Introduction} \label{Introduction}
Both Abelian \cite{buscher:ssbfe} and non-Abelian
\cite{delaossa:1992vc} T--duality assume that an "adapted"
coordinate system consisting of group \coor s and spectators in the
target space of sigma model is given.
However, such system is not unique.

Let us give very simple example for the Euclidean metric
$\eta=\rm{diag}\,(1,1)$ in $\real^2\ni(x,y)$. It is invariant with respect to
one-dimensional isometry group $y\mapsto y+\alpha$ that can be used
for dualization. Of course, the adapted \coor s $(s,g)$ in this case
are $x=s,y=g$ but also
\begin{equation} x=s,\ y=g+\xi(s)\end{equation} where $\xi(s)$ is arbitrary. The Euclidean metric in this \coor s acquires
the form
\begin{equation}\label{eucl in 2dim}
    F_\xi(s,g)=\left(\begin{array}{cc}1+\xi'(s)^2&\xi'(s)\\ \xi'(s)&1\\ \end{array}\right)
\end{equation}
and the dual tensor is
\begin{equation}\label{dual eucl in 2dim}
   \wh F_\xi(s,\ghat)=\left(\begin{array}{cc}1&-\xi'(s)\\ \xi'(s)&1\\
   \end{array}\right).
\end{equation}Evidently, there is no coordinate \tfn{} in the dual
space that would transform $\wh F_\xi(s,\ghat)$ to (symmetric) $\wh
F_{const}(s,\ghat)=\eta$.

Therefore, it is interesting and generally important %In this paper we are going
to investigate how the choice of adapted \coor s, i.e. the choice of
function $\xi(s)$ influences dualization in more complicated
cases.

T--dualities can be understood in the framework of the Poisson--Lie
T--duality introduced in 1995 by C. Klim\v c\'{\i}k and P. \v Severa
using the concept of the \dd{} \cite{klise,klim:proc}. An
important point for construction of dual metrics and torsion
potentials in this framework is whether the group of isometries that
induces the duality acts transitively and freely on the
(pseudo)-Riemannian manifold, in other words, if the manifold is
diffeomorphic to the isometry group. If not, it is necessary to
foliate the manifold by orbits of the isometry group and perform
dualization by the \dd{} method on each of them separately. The
orbits are then numerated by one or more parameters called
spectators since they do not participate in the dualization.

Examples in Refs. \cite{unge:pltp} and \cite{hlapevoj}
choose a specific introduction of spectator $t$ into space-time
metric invariant under the isometry group containing dilations and
shifts of the space and derive corresponding duals. We are going to
show that introduction of spectators is ambiguous and this ambiguity
may influence dualization. The ambiguity follows from
identifications of the Lie group of isometries with its orbits and
as such it is not unique, namely, it depends on the location of the
group unit in the orbit. This ambiguity then appears in the dual
metric and torsion potential as well.

The main question that we want to answer in this paper is whether
these ambiguities are essential for the dual model or if they can be
understood as coordinate effect. In other words, we search
for transformations of coordinates in the dual space that transform
between components of the dual metrics and torsion potentials
obtained from various assignments of the group unit to the invariant
manifolds numerated by the spectators.

In Ref. \cite{delaossa:1992vc} where the dualization is
performed by gauging the non-Abelian symmetry of initial action,
similar ambiguity appears as choice of gauge. It is claimed there
that "Obviously different gauge choices will not give different dual
theories, they will give the same theory differing by a coordinate
\tfn." Goal of this paper is to confirm this conjecture in the
framework of the \dd{} method where the dual tensor field is
obtained explicitly, and mainly, give the procedure for finding the
coordinate \tfn s.

The paper is structured as follows. In Sec.~\ref{sec2}, we give
elements of dualization in the \dd{} framework and present the
appearance of ambiguities. In Sec.~\ref{sec3}, we look for the
\coor{} \tfn s that should eliminate the ambiguities. All steps are
illustrated by dualization of three-dimensional flat metric
expressed in non-adapted \coor s.

\section{T--dualities of \sm s with Spectators}\label{sec2}
We are going to deal with non-Abelian T--dualities of
non-linear sigma models given by the action
\begin{equation}\label{sigm1} S_{\cf}[\phi]=\int_\Omega {\cal L}\,d\tau d\sigma
=\int_\Omega \bar{\cal L}\,d\xi_+ d\xi_- \end{equation} where
$\xi_\pm=\tau\pm\sigma \in \Omega \subset \real^2 $,
\begin{equation} {\cal L}=\half{\cal G}_{\mu\nu}(\phi)\left(\partial_\tau\phi^{\mu}
\partial_\tau \phi^\nu - \partial_\sigma\phi^{\mu}
\partial_\sigma \phi^\nu \right)
+{\cal B}_{\mu\nu}(\phi)\partial_\tau\phi^{\mu}
\partial_\sigma \phi^\nu,
\end{equation}
and  ${\cal G}_{\mu\nu},\ {\cal B}_{\mu\nu}$
 are components of the symmetric and antisymmetric part of
 a tensor field $\cf$ on a manifold  $M$ so that \begin{equation}\bar{\cal L}=\partial_- \phi^{\mu}\cf_{\mu\nu}(\phi)
\partial_+ \phi^\nu=\half{\cal L},\ \ \cf_{\mu\nu}={\cal G}_{\mu\nu}+{\cal
B}_{\mu\nu}.
\end{equation}The
 functions
  $ \phi^\mu:\ \Omega  \rightarrow \real,\
\mu=1,2,\ldots,{\dim}\,M$  are obtained by the composition
$\phi^\mu=x^\mu\circ \phi $ of a map
 $\phi:\Omega\rightarrow M$ and components of a coordinate map $x$ on an open set of  $M$.

 Equations of motion that follow from this action are
 \begin{equation}\label{eqmsm}
  (\cf_{\mu\nu}+\cf_{\nu\mu})\,\partial_{+}\partial_{-}\phi^{\nu} +
  (\cf_{\mu\nu,\beta} + \cf_{\beta\mu,\nu} - \cf_{\beta\nu,\mu})\,
   \partial_{-}\phi^{\beta}\partial_{+}\phi^{\nu} = 0.
 \end{equation}

If $G$ is a non-trivial Lie group of isometries of the tensor field
$\cf$, there is an algebra generated by independent Killing
vector fields $K_a,\ a=1,\ldots,{\dim}\,G$ satisfying
\begin{equation}       \label{killeq}
  (\mathfrak{L}_{K_a}\cf)_{\mu\nu} =0,
 \end{equation}  which is the condition for dualizability of
 $\cf$ where $\mathfrak{L}$ denotes the Lie derivative.

Here we shall focus on the case, when ${\dim}\,G\,<{\dim}\,M$ so
that we can not identify $G\approx M$ (atomic duality).
Nevertheless, we shall assume that the group of isometries acts
freely and transitively on sub-manifolds of $M$ invariant under the
isometry group so that we can identify them with orbits of
$G$. Let us summarize main points of construction of dual models by
the \dd{} method in this case.

\subsection{Invariant sub-manifolds and adapted coordinates}
Before the dualization procedure is started, we have to construct the
invariant sub-manifolds $\Sigma$ of $M$. They are implicitly given by
functions
$\Phi(x^\mu)$
satisfying linear partial differential
equations\begin{equation}\label{invariantconds}
    K_a\Phi=0,\ \ a=1,\ldots,{\dim}\,G.
\end{equation}
As the number of \eqn s is less than the number of independent
variables, we get $S={\dim}\,M\,-{\dim}\,G$ independent solutions
that define the invariant sub-manifolds $\Sigma(s)$ as
\begin{equation}\Phi^\delta(x^\mu)=s^\delta,\ \
 \delta=1,\ldots,S.\end{equation}
{Assuming free action} of the isometry group, we can identify each
of the invariant sub-manifolds with the isometry group and Killing
vectors with left-invariant fields of the group. The latter
identification provides us with transformation to special -- adapted
-- coordinates on $M$ \begin{equation}\label{adapted}
x'^\mu=\{s^\delta,g^a\},\ \
 \delta=1,\ldots,S,\ \ a=1,\ldots,{\dim}\,G,
\end{equation} part of which numerate
the sub-manifolds and the other parametrize group elements by
\begin{equation}g=\e^{g^{1}T_{1}}\e^{g^{2}T_{2}}\ldots\, \e^{g^{\dim G}T_{\dim G}}\end{equation}
where $T_a$ form the basis of the Lie algebra of the group G.

The left-invariant vector fields $V(x')$ are extended to $M$ so
that
\begin{equation}V^\delta(x')=0,\ \delta=1,\ldots,S\end{equation} and \eqn s that
determine \tfn s  to the adapted \coor s are then
\begin{equation}\label{KtoL}
    K_a^\mu(x)=\frac{\partial x^\mu}{\partial x'^\nu}V_a^\nu(x'), \
    a=1,\ldots,{\dim}\,G
\end{equation}where $V_a$ are independent left-invariant fields that
commute in the same way as the corresponding Killing vectors. Number
of equations (\ref{KtoL}) is less than number of sought functions
$X^\mu(x')=x^\mu$, therefore solution will depend on "integration
\fn s"
of spectators $\xi^\mu(s^\delta)$
\begin{equation}\label{xmuXmu}
x^\mu=X^\mu_\xi(x')=X^\mu_\xi(s,g).
\end{equation} As Killing vectors are transformed to left-invariant
vector fields of the group $G$, \coor{} \tfn s (\ref{xmuXmu})
can be understood as right action of the group $G$ in the invariant
sub-manifold $\Sigma(s)$ and
\begin{equation}\label{rgaction}
x^\mu= (u_\xi(s)\vartriangleleft g)^\mu
\end{equation} where $u_\xi(s)$ are
points of sub-manifolds $\Sigma(s)$ that correspond to the unit
element $u\in G$. It means that  functions $\xi^\mu$ must
satisfy condition\begin{equation}\label{xionSigma}
\Phi^\delta(\xi^\mu)=s^\delta,\ \
 \delta=1,\ldots,S.
\end{equation}

For application of dualization procedure, components of the
tensor field $\cf$ must be expressed in the adapted coordinates
(\ref{adapted}) as
\begin{equation}\label{FtoF}
    F_{\kappa\lambda}(s,g)=\frac{\partial x^\mu}{\partial x'^\kappa}
    \frac{\partial x^\nu}{\partial
    x'^\lambda}\cf_{\mu\nu}(x).
\end{equation}It is clear that components of the tensor field $\cf$
in adapted \coor s will depend on functions $\xi^\mu$ since the
\tfn{} matrix $\frac{\partial x^\mu}{\partial x'^\kappa}$ does. The
\eqn s (\ref{killeq}) take then the form
\begin{equation}       \label{kseq0}
  (\mathfrak{L}_{V_{a}}F)_{\mu\nu} =0,
 \end{equation}
 which is the
condition for non-Abelian dualizability of the \sm{} \cite{klise}.
\subsubsection{Example}The simplest example of non-Abelian T--duality with one spectator
can be given for three-dimensional metrics invariant under
two-dimensional non-Abelian group. Let us have pseudo-Riemannian flat
metric $\eta= \rm{diag}\,(-1,1,1)$ in $\real^3\ni(t,y,z)$ invariant
under subgroup $G_2$ of inhomogeneous Lorentz group ISO(1,2)
generated by the Killing vector fields \cite{PWZ}
\begin{equation}\label{kils2}
    K_1=-z\,\partial_t-t\,\partial_z,\
    K_2=-y\,\partial_t-(t+z)\,\partial_y+y\,\partial_z
\end{equation}commuting as \begin{equation} [K_1,K_2]=-K_2.\end{equation} The sub-manifolds of $\real^3$ invariant
under this subgroup are given by condition \begin{equation}\Phi(t,y,z)=t^2-y^2-z^2=const\end{equation} where the value of the constant
depend on the sub-manifold and can be taken as  spectator $s$. The
Killing vector fields are transformed to left-invariant vector
fields
\begin{equation}\label{lvers2} V_1=\partial_1+g_2\partial_2,\
    V_2=\partial_2
\end{equation}of the group \begin{equation}\label{S2}
    G_2=\{g\in\real^2,\ g\cdot g':=(g_1+g'_1,\e^{g'_1}g_2+g'_2)\}
\end{equation}
by  transformation $(t,y,z)\mapsto(s,g_1,g_2)$
\begin{eqnarray}\label{xtosgex1}
\nonumber
  t &=& \frac{1}{2} \left(\e^{-g_1} \left(g_2^2+1\right) \xi_1(s)+\e^{g_1} \xi_3(s)-2
   g_2\,\xi_2(s)\right), \\
  y &=& \xi_2(s)-\e^{-g_1} g_2\,\xi_1(s), \\
\nonumber  z &=& -\frac{1}{2}\left(
   \e^{-g_1} \left(g_2^2-1\right) \xi_1(s)+\e^{g_1}
   \xi_3(s)-2g_2\,\xi_2(s)\right).
\end{eqnarray} The \tfn{} depends on functions $\xi_j(s)$ that fix the positions
\begin{equation}\label{exmplcurve}
u_\xi(s)=\left(\frac{\xi_1(s)+\xi_3(s)}{2},\xi_2(s),\frac{\xi_1(s)-\xi_3(s)}{2}\right)
\end{equation}
of the unit of $G_2$ in the invariant manifolds $\Sigma(s)$. They are
arbitrary up to condition \begin{equation}\label{exmplcond}
\xi_1(s)\,\xi_3(s)-\xi_2(s)^2=s
\end{equation} that follows from the requirement
$t^2-y^2-z^2=s.$ The \tfn{} (\ref{xtosgex1}) is transition from the
space-time coordinates of $\real^3$ to the group coordinates of $G_2$
and spectator $s$ numerating the invariant sub-manifolds. It is easy,
even though tedious, to check that this \tfn{} is right action
(\ref{rgaction}) of $G_2$ on $\real ^3$. Moreover, for $\xi_1(s)\neq
0$, the Killing vectors are independent and the curve
(\ref{exmplcurve}) in $\real^3$ is transversal to the invariant
sub-manifolds of $\real^3$.

Components $F_{\kappa\lambda}(s,g_1,g_2)$ of the flat metric $\eta$
in the adapted coordinates are rather complicated and depend heavily
on functions $\xi_j(s)$ and their derivatives $\xi_j'(s)$ as
\begin{equation}\label{exmplF}
F_\xi(s,g) =\left(
\begin{array}{ccc}
 \frac{\left(\xi_2^2+s\right) \xi_
   1'^2-\xi_1 \left(2 \xi_2 \xi_
   2'+1\right) \xi_1'+\xi_1^2 \xi_
   2'^2}{\xi_1^2} & \frac{\xi_1 \left(2
   \xi_2 \xi_2'+1\right)-2 \left(\xi_
   2^2+s\right) \xi_1'}{2 \xi_1} &
   \e^{-g_1} \left(\xi_2 \xi_
   1'-\xi_1 \xi_2'\right) \\
 \frac{\xi_1 \left(2 \xi_2 \xi_
   2'+1\right)-2 \left(\xi_2^2+s\right) \xi_
   1'}{2 \xi_1} & \xi_2^2+s &
   -\e^{-g_1} \xi_1 \xi_2 \\
 \e^{-g_1} \left(\xi_2 \xi_1'-\xi_
   1 \xi_2'\right) & -\e^{-g_1} \xi_
   1 \xi_2 & \e^{-2 g_1} \xi_1^2 \\
\end{array}
\right).
\end{equation}
Here we have put
\begin{equation} \xi_3(s)=\frac{s+\xi_2(s)^2}{\xi_1(s)}\end{equation} to satisfy (\ref{exmplcond}).
\subsection{Dualization}
Non-Abelian T--duality is a special case of the \pltd y \cite{klise,klim:proc} formulated in the framework of the \dd{}  -- a Lie
group whose Lie algebra $\cd$ admits a decomposition
$\cd=\cg\dotplus\wh\cg$ into a pair of sub-algebras maximally
isotropic with respect to a symmetric ad-invariant non-degenerate
bi-linear form $\langle\, .\,,.\,\rangle_\cd $.

\pltd y can be applied to models with tensor fields satisfying
\cond{}
\begin{equation}       \label{kseq}
  (\mathfrak{L}_{V_{a}}F)_{\mu\nu} =
   F_{\mu\kappa}V^{\kappa}_{b}\wh{f}^{bc}_{a} %krat
   V^{\lambda}_{c}F_{\lambda\nu}
 \end{equation}
where $V_b^\kappa$ are components of the left-invariant vector
fields $V_b$ generating a Lie group $G$ and $\wh{f}^{bc}_{a}$ are
 structure coefficients of a "dual" Lie group $\wh G$ of the same dimension.
In case of non-Abelian T--duality $\wh{f}^{bc}_{a}=0$ so that the dual
group is Abelian. Self-consistency of the
 condition (\ref{kseq}) restricts the structure coefficients in such way
 that $G$ and $\wh G$ can be interpreted
 as subgroups defining the \dd{} $D\equiv(G|\wh G)$.

Components of tensor field $\cf$ satisfying the \cond{}
(\ref{kseq0}) %with $\wh{f}^{bc}_{a}=0$
can be written as
\begin{equation}\label{met}
        (F_\xi)_{\mu\nu}(s,g)=e_{\mu}^{j}(g)(E_\xi(s))_{jk}e_{\nu}^{k}(g)
    \end{equation}where the matrix $E_\xi(s)=F_\xi(s,0)$,
\begin{equation}
e_{\mu}^{j}(g)=\left(\begin{array}{cc} \unit_S & 0 \\ 0
&e_{\mu}^{a}(g)
     \end{array}\right)\end{equation}
and $e_{\mu}^{a}(g)$ are components of right-invariant forms
$(dg)g^{-1}$
in adapted coordinates.

Components of the dual tensor $\wh\cf$ obtained by the
non-Abelian T--duality \tfn{} are \cite{klise}
\begin{equation}\label{Fghat2} \widehat F_\xi (s,\ghat)=(\wh E_\xi(s)^{-1}+\wh\Pi(\ghat))^{-1} \end{equation}
where $\widehat E_\xi(s)$ is \begin{equation}\label {Extil}
  \wh E_\xi(s)=\big(A+E_\xi(s)\cdot{B}
    \big)^{-1}\big(B+E_\xi(s)\cdot A\big),
\end{equation}
\begin{equation}\label{matsAB}
{A}=\left(\begin{array}{cc} \unit_S & 0 \\ 0 & \bold O_G
     \end{array}\right), \quad
{B}=\left(\begin{array}{cc}  \bold O_S & 0 \\ 0 & \unit_G
     \end{array}\right).\end{equation}
Matrices $\unit_G,\unit_S $ and $\bold O_G,\bold O_S$ are unit and
zero matrices of ${\dim}\,G$ and $({\dim}\,M-{\dim}\,G)$,
\begin{equation} \label{Pihat2}
  \widehat\Pi(\hat g)=\left(\begin{array}{cc}  \bold O_S & 0 \\ 0 & -{f_{cd}}^{b}\ghat_b%\wh b(\hat g)
  \end{array}\right)
\end{equation}%components of the matrix  $\wh b$ are $\wh b_{ab}(\hat g)=-{f_{ab}}^{c}\ghat_c $
where ${f_{cd}}^{b}$ are structure coefficients of the Lie algebra
of the group  $G$ and $\ghat_b$ are coordinates of the Abelian group
$\wh G$. As both factors in the definition of $\wh E_\xi(s)$
(\ref {Extil}) must have non-vanishing determinants, we get from
(\ref{Extil}), (\ref{matsAB}) conditions
\begin{equation}\label{condsAB}
 \det\,  (A+E_\xi(s)\cdot{B})\neq 0,\ \  \det\, (B+E_\xi(s)\cdot A)\neq 0
\end{equation}
that further restrict the \fn s $\xi^\mu$.

Formula (\ref{Fghat2}) for the non-Abelian T--duality of \sm s with
spectators follow from the \pltd y formulated in the framework of
the \dd{} \cite{klise,klim:proc,hlapevoj}. By this formula we get
dual tensors whose components
will depend on functions $\xi^\mu$ or $X_\xi$
mapping for fixed $s$ from the group $G$ to the invariant manifold
$\Sigma(s)$.

\subsubsection{Dual tensor field -- example continues}
From the expression (\ref{met}) we can get the matrix $E(s)=
E_\xi(s)$ by setting $g_1=0$ in (\ref{exmplF}). Formula
(\ref{Extil}) then yields
\begin{equation}\label{exmplExtil}
\wh E_\xi(s)=  \left(
\begin{array}{ccc}
 -\frac{1}{4 s} & \frac{\xi_1'(s)}{\xi_
   1(s)}-\frac{1}{2 s} & -\frac{\xi_2(s)-2 s\, \xi_
   2'(s)}{2 s\, \xi_1(s)} \\
 \frac{1}{2 s}-\frac{\xi_1'(s)}{\xi_1(s)} &
   \frac{1}{s} & \frac{\xi_2(s)}{s\, \xi_1(s)} \\
 \frac{\xi_2(s)-2 s\, \xi_2'(s)}{2 s\, \xi_
   1(s)} & \frac{\xi_2(s)}{s\, \xi_1(s)} &
   \frac{\xi_2(s)^2+s}{s\, \xi_1(s)^2} \\
\end{array}
\right)
\end{equation}
and applying the formula (\ref{Fghat2}) with \begin{equation}\wh\Pi(\ghat)=\left(
\begin{array}{ccc}
 0 & 0 & 0 \\
 0 & 0 & \ghat_2 \\
 0 & -\ghat_2 & 0 \\
\end{array}
\right),\end{equation} we get the dual tensor field in adapted coordinates
$(s,\ghat_1,\ghat_2)$ with non-vanishing torsion and scalar
curvature
\begin{equation} \wh R=\frac{4\, \xi_1(s)^2\left(11\, \ghat_2^2\, -
3\, s\,\xi_1(s)^2\right)}{\left(\ghat_2^2+s\,
\xi_1(s)^2\right)^2}.\end{equation} Unfortunately, $\wh F_\xi(s,\ghat)$ is too
extensive to display it here.

\section{Change of Functions $\xi^\mu$ versus Coordinate Transformations
}\label{sec3} It is clear from the above that components of both
tensor fields $\cf$ and $\wh \cf$ in adapted \coor s may depend on
functions $\xi^\mu(s)$. The question is whether we can get rid off
this dependence by a coordinate \tfn{} in both initial and dual
tensors. More precisely, we search for \tfn{} of group coordinates
$g^a=\gamma^a(s,\gtil)$ and $\ghat^a=\hat\gamma^a(s,\gbar)$  that
would be equivalent to the change of \fn s
$\xi^\mu(s)\mapsto\xitil^\mu(s)$, i.e.\begin{equation}\gamma^a:\
 e(g)E_\xi(s)e(g)^T\mapsto e(\gtil)E_{\xitil}(s)e(\gtil)^T\end{equation} and
\begin{equation}\hat\gamma^a :\  (\wh E_\xi(s)^{-1}+\wh\Pi(\ghat))^{-1}\mapsto (\wh E_{\xitil}(s)^{-1}+\wh\Pi(\gbar))^{-1}.\end{equation}

It means that we look for \fn s
$\Gamma(s,\gtil)=(s^\delta,\gamma^a(s,\gtil))$ and
$\wh\Gamma(s,\gbar)=(s^\delta,\hat\gamma^a(s,\gbar))$ such that
\begin{equation}\label{xichange}
(F_{\xitil})_{\mu\nu}(s,\gtil)=\frac{\partial
\Gamma^\kappa}{\partial \tilde{x}'^\mu}\frac{\partial
\Gamma^\lambda}{\partial
\tilde{x}'^\nu}(F_{\xi}(\Gamma))_{\kappa\lambda}\end{equation} and
\begin{equation}\label{xichangehat} (\wh F_{\xitil})_{\mu\nu}(s,\gbar)=\frac{\partial
\wh\Gamma^\kappa}{\partial \bar{x}'^\mu}\frac{\partial
\wh\Gamma^\lambda}{\partial \bar{x}'^\nu}(\wh
F_{\xi}(\wh\Gamma))_{\kappa\lambda}.
\end{equation}The answer to
the question above is partially positive in the sense that both \fn
s $\gamma^a$  and $\hat\gamma^a$ are given implicitly in general,
and many examples show that
instead of (\ref{xichangehat}) we always get (see the example in the
Introduction)
\begin{equation}\label{xichangehat2}
(\wh F_{\xitil})_{\mu\nu}
=\frac{\partial \wh\Gamma^\kappa}{\partial \bar
x'^\mu}\frac{\partial \wh\Gamma^\lambda}{\partial \bar x'^\nu}(\wh
F_{\xi})_{\kappa\lambda}
+ \Omega
,\ \ d\Omega =0.
\end{equation}  How to find functions $\Gamma$ and $\wh\Gamma$?
\subsection{Transformation of \coor s on the initial manifold}
Let us start with the case where the manifold $M$ has structure of
Lie group but dualization is performed only with respect to a
subgroup $G\subset M$. The right action of the subgroup
 $G$ on the location of the group unit
$u_\xi(s)$ in the sub-manifold $\Sigma(s)\subset M$
is then realized by  group multiplication in $M$ as
\begin{equation} x = u_\xi(s)\cdot g,\ \ g\in G. \end{equation}
From this we get the transformation (\ref{xmuXmu}) of manifold
coordinates $x^\mu$ to the group parameters and spectators $x'^\mu$.
Changing the location of the unit element on the orbit $\Sigma(s)$
to $u_{\tilde\xi}\,(s)$, the same point $x\in\Sigma(s)$ is obtained
by a different group element
\begin{equation} x = u_{\xitil}\,(s)\cdot\tilde g,\ \ \tilde g\in G. \end{equation}
Comparing these two \eqn s, we get corresponding change of subgroup
elements induced by the change of functions $\xi(s)$
\begin{equation}\label{grouptfn1}
    g=u_\xi(s)\-1\cdot u_{\tilde\xi}\,(s)\cdot\tilde g
\end{equation}
and consequently the \tfn{} of the group parameters
$g^a=\gamma^a(s,\gtil)$ corresponding to the change of \fn s
$\xi\rightarrow\xitil$.

If the manifold has not structure of the Lie group and the right
action of the group G is given by a more general map
(\ref{xmuXmu}), it is not difficult to deduce that transformation of
the group
 $G$ \coor s $g^a=\gamma^a(s,\gtil)$ will be obtained from requirement
\begin{equation}\label{gammaimplicit}
X^\mu_\xi(s^\delta,g^{a}) =x^\mu=X^\mu_{\xitil}(s^\delta,\gtil^b).
\end{equation}
However, in this case differently from (\ref{grouptfn1}), \tfn{}
$g^{a}=\gamma^a(s,\gtil)$ is given by the \eqn{}
(\ref{gammaimplicit}) only implicitly and must be calculated for
each $X^\mu$ separately. E.g.,
{the \tfn{} of group coordinates following from (\ref{xtosgex1}) is
\begin{equation}\label{exmplgamma}
   {g_1}= {\tilde g_1}+\log\frac{ \xi_1(s)}{\tilde\xi_1(s)},\ \
    {g_2}= {\tilde g_2}+ \e^{{\tilde g_1}}\frac{ {{\xi_2}(s)-{\tilde\xi_2}(s)}}
    {\tilde\xi_1(s)}.
\end{equation}
}\subsection{Transformation of \coor s on the dual manifold}
More interesting and more important is  transformation of components
of the dual tensor. Its form  $\ghat^a=\hat\gamma^a(s,\gbar)$ can be
obtained from dual decomposition of elements of the \dd{}
\begin{equation}\label{ghtil=gtilh}
    l=g\cdot\htil=\ghat\cdot h, \ \ g,h\in G,\ \htil,\ghat\in\wh G.
\end{equation}This formula can be used for solution
of the \eqn s of motion of the \sm{} in the dual tensor field
\cite{hlape:ppw} but here we will use it for finding the \coor{}
\tfn{} in the dual space.

To find the dual decomposition for given $g,\htil$ is rather
complicated problem in general but its solution simplifies
substantially in case of (non)-Abelian T--duality where the dual group
$\wh G$ is Abelian. In this case we can use a representation $r$ {of
an element of} the semi-Abelian \dd{} in terms of block matrices
$(\dim G+1)\times (\dim G+1)$ with
\begin{equation}\label{Drep}
   r(g)= \left(
\begin{array}{cc}
Ad\, g & 0 \\
 0& 1
\end{array}
\right),\ \ \   r(\htil)= \left(
\begin{array}{cc}
\unit_G& 0 \\
 \vec v(\wt h)& 1
\end{array}
\right)
\end{equation}
where $\vec v(\htil)=(\htil^1,\ldots,\htil^{\dim G}),\ h^j$ being
parameters of the (Abelian) group element \begin{equation}\htil=\e^{\htil_{1}\hat
T^{1}}\e^{\htil_{2}\hat T^{2}}\cdots \e^{\htil_{\dim  G}\hat T^{\dim
G}}. \end{equation}From the equation (\ref{ghtil=gtilh}) we then get
\begin{equation}\label{replgh}
   r(l)=r(g\htil)= \left(
\begin{array}{cc}
Ad\, g & 0 \\
 v(\htil)& 1
\end{array}
\right) =r(\ghat  h)= \left(
\begin{array}{cc}
Ad\, h & 0 \\
 v(\ghat)\cdot(Ad\, h )& 1
\end{array}
\right). \end{equation} If the adjoint representation of the Lie
algebra $\mathfrak g$ is faithful then the representation $r$ of the
\dd {} is faithful as well and the relation (\ref{replgh})
immediately gives $g=h$. If not, we can use formula
\begin{equation}\label{BHC}
    \e^{A}\e^{B}=\e^{(Ad\,A)\, B}\,\e^A%\e^{\exp(adA) B}\e^A
\end{equation}
to permute the elements of $G$ and $\wh G$ in (\ref{ghtil=gtilh})
and again we get $g=h$. From the decomposition (\ref{ghtil=gtilh})
we then get \begin{equation}\label{htil} \htil=g\-1\cdot
\ghat\cdot g.\end{equation} Similarly for $\gtil, \gbar$ we get %will write
\begin{equation}\label{ghtil=gtilh2}
    \htil=\gtil\-1\cdot \gbar\cdot\gtil, \ \ \gtil\in G,\ \htil,\gbar\in\wh
    G.
\end{equation}
Comparing (\ref{htil}) and (\ref{ghtil=gtilh2}), we get transition
$\ghat \mapsto\gbar$
\begin{equation}\label{gtiltogbar}
    \ghat=g\cdot \gtil\-1\cdot\gbar\cdot \gtil\cdot g\-1
\end{equation} corresponding to $\xi\mapsto\xitil$.

For manifolds with group structure one can see from
(\ref{grouptfn1}) that \begin{equation}\gtil\cdot g\-1=u_{\xitil}(s)\-1\cdot
u_{\xi}(s)\end{equation} so that the relation between $\gbar$ and $\ghat$
depends only on $\xi^\mu(s)$ and $\xitil^\mu(s)$. By expressing
(\ref{gtiltogbar}) in group parameters, we get the \fn{}
$\hat\gamma^a(s,\gbar)$ that transform dual tensor fields $\wh
F_\xi$ to $\wh F_{\xitil}$ by (\ref{xichangehat2}). For general case
where relation between $g$ and $\ghat$ is given implicitly by
(\ref{gammaimplicit}), \fn{} $\hat\gamma^a(s,\gbar)$ again depends
only on $\xi^\mu(s)$ and $\xitil^\mu(s)$ since from (\ref{rgaction})
we get
\begin{equation}\label{gtilgin}
    u_\xi(s)=u_{\tilde\xi}\,(s)\vartriangleleft(\gtil\cdot g\-1)
\end{equation}so that solution of this \eqn{} for $\gtil\cdot g\-1$ must
depend only on $\xi^\mu(s)$ and $\xitil^\mu(s)$. E.g., the
\tfn{} of the dual group \coor s in the example given above is
\begin{equation}\ghat_1=\gbar_1+{\gbar_2}\frac{ {\xitil_2}(s) -
   \xi_2(s)}{\xitil_1(s)},\ \
    \ghat_2={\gbar_2}\frac{\xi_1(s)}{\xitil_1(s)}\end{equation} and \begin{equation} \Omega=
\begin{array}{c}
 ds\wedge d\gbar_1\left(\frac{\xi_1'(s)}{\xi_1(s)}-\frac{\xitil_1'(s)}{\xitil_1(s)}\right)
 +ds\wedge d\gbar_2\frac{\xi_1'(s)\left(\xitil_2(s)-\xi_2(s)\right)+\xi_1(s) \left(\xi_2'(s)-
   \xitil_2'(s)\right)}{\xi_1(s) \xitil_1(s)}
\end{array}.
\end{equation}

\section{Conclusion} We have shown that both
initial and dual tensors $\cf$ and $\wh\cf$ defining \sm s expressed
in adapted \coor s may depend on functions of spectators
$\xi^\mu$
 that are arbitrary up to conditions (\ref{xionSigma}) and
(\ref{condsAB}). Corresponding choices of sub-manifolds (curves,
surfaces, etc.) representing the choices of location of the isometry
group unit in the invariant sub-manifolds can be to large extent
compensated by \coor{} \tfn{}. We have given the formulas for the
corresponding \coor{} \tfn s.

Let us note that when deriving transition
$\ghat^{a}\mapsto\gbar^{a}$ transforming components of dual tensor
field $\wh \cf$ in various adapted coordinates, we have
substantially used the fact that the \dd{} is semi-Abelian.
Therefore, it is not clear whether this \tfn{} can be found for
general \pltd y or \pltp y with
spectators and, especially, how to find $\hat\gamma^a$ in these general cases.

\section*{Acknowledgment}

This work was supported by the Grant Agency of the Czech Technical University in Prague, grant No. SGS16/239/OHK4/3T/14.

\end{document}